\documentclass[prl,twocolumn,showpacs,preprintnumbers,amsmath,amssymb,superscriptaddress]{revtex4-1}

\usepackage{graphicx}
\usepackage{dcolumn}
\usepackage{bm}
\usepackage{epsfig}
\usepackage{amsmath}
\usepackage{amssymb}
\usepackage{color}
\usepackage{bbm}
\usepackage[caption=false]{subfig}
\usepackage{placeins}
\usepackage{soul}
\usepackage{braket}
\usepackage{float}
\usepackage{multirow}
\usepackage{diagbox}

\definecolor{mybo}{cmyk}{0,0,0.10,0}

\usepackage[colorlinks=true,citecolor=blue,linkcolor=blue,urlcolor=blue]{hyperref}
\usepackage{cancel}

\def\equationautorefname~#1\null{%
	Eq.~#1\null
}
\def\figureautorefname~#1\null{%
	Fig.~#1\null
}

\begin{document}

\title{Experimental observation of parity-symmetry-protected phenomena in the quantum Rabi model with a trapped ion}

\author{Xingyu Zhao}
\email{These authors contributed equally to this work.}
\affiliation{CAS Key Laboratory of Microscale Magnetic Resonance and School of Physical Sciences, University of Science and Technology of China, Hefei 230026, China}
\affiliation{Anhui Province Key Laboratory of Scientific Instrument Development and Application, University of Science and Technology of China, Hefei 230026, China}
\affiliation{Hefei National Laboratory, University of Science and Technology of China, Hefei 230088, China}

\author{Qian Bin}
\email{These authors contributed equally to this work.}
\affiliation{College of Physics, Sichuan University, Chengdu 610065, China}
\affiliation{School of Physics and Institute for Quantum Science and Engineering, Huazhong University of Science and Technology, and Wuhan Institute of Quantum Technology, Wuhan 430074, China}

\author{Waner Hou}
\affiliation{CAS Key Laboratory of Microscale Magnetic Resonance and School of Physical Sciences, University of Science and Technology of China, Hefei 230026, China}
\affiliation{Anhui Province Key Laboratory of Scientific Instrument Development and Application, University of Science and Technology of China, Hefei 230026, China}

\author{Yi Li}
\affiliation{CAS Key Laboratory of Microscale Magnetic Resonance and School of Physical Sciences, University of Science and Technology of China, Hefei 230026, China}
\affiliation{Anhui Province Key Laboratory of Scientific Instrument Development and Application, University of Science and Technology of China, Hefei 230026, China}
\affiliation{Hefei National Laboratory, University of Science and Technology of China, Hefei 230088, China}

\author{Yue Li}
\affiliation{CAS Key Laboratory of Microscale Magnetic Resonance and School of Physical Sciences, University of Science and Technology of China, Hefei 230026, China}
\affiliation{Anhui Province Key Laboratory of Scientific Instrument Development and Application, University of Science and Technology of China, Hefei 230026, China}

\author{Yiheng Lin} 
\email{yiheng@ustc.edu.cn}
\affiliation{CAS Key Laboratory of Microscale Magnetic Resonance and School of Physical Sciences, University of Science and Technology of China, Hefei 230026, China}
\affiliation{Anhui Province Key Laboratory of Scientific Instrument Development and Application, University of Science and Technology of China, Hefei 230026, China}
\affiliation{Hefei National Laboratory, University of Science and Technology of China, Hefei 230088, China}

\author{Xin-You L\"{u}}
\email{xinyoulu@hust.edu.cn}
\affiliation{School of Physics and Institute for Quantum Science and Engineering, Huazhong University of Science and Technology, and Wuhan Institute of Quantum Technology, Wuhan 430074, China}

\author{Jiangfeng Du} 
\affiliation{CAS Key Laboratory of Microscale Magnetic Resonance and School of Physical Sciences, University of Science and Technology of China, Hefei 230026, China}
\affiliation{Anhui Province Key Laboratory of Scientific Instrument Development and Application, University of Science and Technology of China, Hefei 230026, China}
\affiliation{Hefei National Laboratory, University of Science and Technology of China, Hefei 230088, China}
\affiliation{Institute of Quantum Sensing and School of Physics, Zhejiang University, Hangzhou 310027, China}

\date{\today}

\begin{abstract}

Symmetry is crucial for gaining insights into the fundamental properties of physical systems, bringing possibilities in studying exotic phenomena such as quantum phase transitions and ground state entanglement.  Here, we experimentally simulate a highly controllable extended quantum Rabi model, capable of tuning into the ultra-strong or deep coupling regime, in a spin-motion-coupled trapped ion. We observe that the phonon driven by such a model with parity symmetry preserved (broken) would experience double (single) excitation in the ultra-strong coupling regime. Quantum phenomena such as strong ground state entanglement and quantum superposition in systems occur with parity symmetry, and these phenomena disappear following the symmetry breaking. We also find sensitive responses for the two-level system entropy and phonon Wigner function in the deep coupling regime, depending on the parameter across the symmetry transition point. This work offers the prospect of exploring symmetry-controlled quantum phenomena and their applications in high-precision quantum technologies.

\end{abstract}
\maketitle

\emph{Introduction}.---Symmetry is fundamental to our understanding and exploration of the physical world, underpinning the formulation of basic principles\,\cite{Weyl1931} and playing a role in practical applications like stability criteria\,\cite{Marsden1999Ratiu}. It has been demonstrated that symmetry helps in obtaining the hydrogen atom spectrum \,\cite{Schiff1955}, studying local hidden variable models\,\cite{Werner1989}, exploring quantum phase transitions\,\cite{S. Sachdev, R. Coldea, K. Baumann}, and topological classification\,\cite{C. -K. Chiu, Gong2018AK}. The presence of certain symmetries can also simplify dynamics calculations by organizing the kinematic space of theories using symmetry groups\,\cite{S. Singh}. 
For example, the quantum Rabi model (QRM), consisting of a bosonic mode coupled to a two-level system (TLS), typically exhibits a parity (or $\mathbb{Z}_2$) symmetry, indicating integrability\,\cite{D. Braak2011} and confinement of dynamical evolution to parity-conserving subspaces\,\cite{J. Casanova2010}. It is a basic model to describe light-matter interaction and has produced fruitful research results in different aspects\,\cite{Qiongtao Xie}. 

Generally, QRM can be divided into several regimes according to the coupling strength. In the weak or strong coupling regime, the rotating wave approximation (RWA) applies, reducing the QRM with discrete $\mathbb{Z}_2$ (parity) symmetry to the analytically solvable Janes-Cummings model\,\cite{Jaynes1963} with continuous $U(1)$ symmetry. In the ultra-strong (or deep-strong coupling) regimes, where the coupling strength reaches ten percent of (or exceeds) the mode frequency, the RWA breaks down and a full QRM with parity symmetry is required to properly describe the dynamical and equilibrium properties\,\cite{Kockum2019, Gu2017, FornDiaz2019}. As the coupling strength surpasses a critical point, the ground state of the QRM undergoes spontaneous parity symmetry breaking, leading to the gradual emergence of doubly degenerate energy levels\,\cite{S. Ashhab}. This corresponds to a quantum phase transition at zero temperature\,\cite{Myung-Joong}, which has been experimentally observed using a single trapped ion\,\cite{Cai2021LZ}. Furthermore, the parity symmetry of the system Hamiltonian can be broken by introducing an additional term that explicitly controls the symmetry. The absence of parity symmetry in the Hamiltonian significantly alters the system's behavior, giving rise to parity-sensitive quantum phenomena\,\cite{D. Braak2011, Ma2015Law, Garziano2016MS, Bin2018LY, F. Deppe, Ridolfo2012LS, Malekakhlagh2019Rodriguez, Bin2021WL}. While experimental probing of symmetry breaking in extended QRM has been achieved in circuit quantum electrodynamics systems\,\cite{F. Deppe, Wang2023RL}, many exotic quantum phenomena closely related to the parity symmetry of the Hamiltonian remain elusive.

In this work, we experimentally demonstrate a direct method\,\cite{Bin2021WL} for probing the parity symmetry of the extended QRM and exploring parity-symmetry-controlled quantum phenomena using a trapped ion. We show that the peak shape of bosonic mode excitation can serve as a sensitive probe for determining the system's parity symmetry. Compared to the previous work\,\cite{Wang2023RL}, which probed parity symmetry breaking in the extended quantum Rabi model (QRM) using an ancillary qubit and restricted observations of parity symmetry signals to the near deep-strong coupling regime, here we determine the system's parity symmetry directly by observing phonon excitation peaks induced by sideband transitions after driving the TLS in a trapped ion. This method eliminates the need for an ancillary mode and is effective in both the ultra-strong and deep-strong coupling regimes.

Moreover, due to the full controllability and tunability of the trapped ion platform~~\cite{Leibfried2003BM, Pedernales2015}, the system in various coupling regimes can be readily simulated.  We experimentally demonstrate that in the symmetry-conserving case, strong ground-state entanglement and quantum superposition emerge when the coupling strength surpasses the critical point. This is evidenced by a large von Neumann entropy and the presence of Schr\"{o}dinger cat states characterized by negative Wigner distributions with well-separated peaks and interference fringes. Importantly, beyond the parameter at the phase transition point, the entropy and Wigner function of the ground state are highly sensitive to the parity symmetry of the system Hamiltonian. The observed strong ground-state entanglement and quantum superposition that arise from ground-state spontaneous symmetry breaking do not occur in the system without parity symmetry.  These results are not only fundamentally interesting but can also inspire the engineering of new symmetry-manipulated quantum devices or extending the coherence time to get long-lasting quantum memories\,\cite{Stassi2018Nori}.

\emph{Model and experimental setup}.---The extended QRM model we consider here is shown in Fig.\,\ref{fig1}(a). Its Hamiltonian is written as ($\hbar=1$)
\begin{align}\label{eq01}
\begin{split}
H_s= \frac{\omega_{\sigma}}{2}\sigma_z+\omega_a a^\dag a+ \lambda(\cos \!\theta\sigma_z-\sin \!\theta \sigma_x)(a + a^\dag) \,,\\
\end{split}
\end{align}
where $a$ ($a^\dag$) represents the annihilation (creation) operator of the bosonic mode with the resonant frequency $\omega_a$, $\sigma_x=\sigma^\dag+\sigma$ ($\sigma=|g\rangle\langle e|$) and $\sigma_z$ are Pauli operators of the TLS with the transition frequency $\omega_{\sigma}$, $\lambda$ is the coupling strength between the TLS and the bosonic mode, and the angle $\theta\in[0,\pi]$ determines the relative contribution of the longitudinal and transverse couplings. 
When $\theta=\pi/2$, the coupling is transverse, and the Hamiltonian in Eq.\,(\ref{eq01}) reduces to the standard Rabi Hamiltonian. This model exhibits parity symmetry, characterized by a well-defined parity operator $\Pi=e^{i\pi \mathbb{N}}$, with $[H_s,\Pi]=0$\,\cite{D. Braak2011,J. Casanova2010}.  Here $\mathbb{N}=a^\dag a+(\sigma_z+1)/2$ represents the total excitation number. The parity eigen-equation $\Pi|p_i\rangle_{\pm}=\pm|p_i\rangle_{\pm}$ ($i=1,2,3\dots$) defines the odd-even parity states, which constitute two independent parity-conserving subspaces $\{|p_i\rangle_{+} \}$ and $\{|p_i\rangle_{-} \}$\,\cite{J. Casanova2010, Bin2021WL}. States with a certain initial parity evolve only within the corresponding subspace, and transitions between the two different parity subspaces are forbidden. When $\theta \neq \pi/2$, the presence of longitudinal coupling breaks the system's parity symmetry, with $[H_s,\Pi]\neq0$. In this case,  transitions between the two different parity subspaces become allowed.  Moreover, it is well known that increasing the coupling strength to a critical point $\lambda=\sqrt{\omega_a\omega_{\sigma}}/2$ can induce a phase transition in the standard QRM
. However, for $\theta\neq\pi/2$, this phase transition does not occur due to the absence of the system parity symmetry, and there is no abrupt change in the average phonon number $\langle a^{\dag} a\rangle$ and state $|e\rangle$ population $(\left<\sigma_z\right>+1)/2$ of the ground state, as shown in Fig.\,\ref{fig1}(c,\,d).

\begin{figure}
\includegraphics[width=8.7cm]{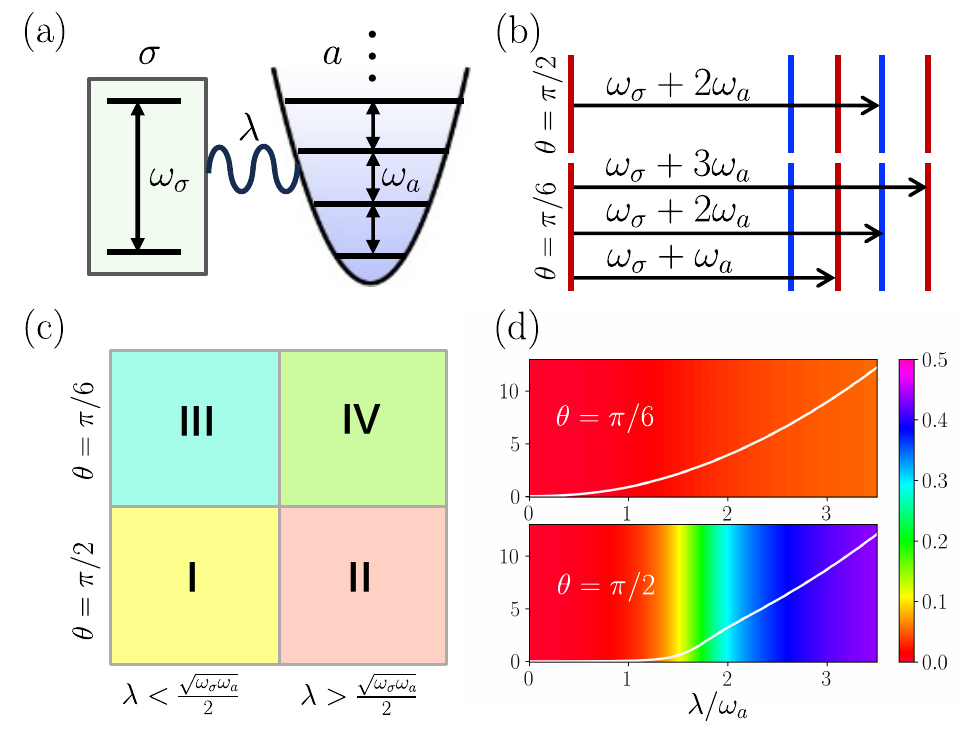}\\
\caption{(a) The quantum Rabi model: a TLS coupled to a bosonic mode, corresponding to the interaction between the electronic state and axial motion of a trapped ion mediated by a laser field. (b) Sideband transitions for $\theta=\pi/2$ and $\theta\neq\pi/2$. The red lines indicate states $|0,g\rangle$, $|1,e\rangle$, $|3,e\rangle$ with parity $P=+1$, while the blue lines indicate states $|0,e\rangle$, $|2,e\rangle$ with parity $P=-1$.  (c,d) Average phonon number $\langle a^\dag a\rangle$ (white lines) and state $|e\rangle$ population $(\left<\sigma_z\right>+1)/2$ (background color) (d) in the ideal ground state in two distinct parameter regimes (c), where $\omega_{\sigma}/\omega_a = 6$.} \label{fig1}
\end{figure}

The experiment is carried out with a single $^{40} {\rm Ca} ^{+}$ ion trapped in a linear Paul trap. The ion's axial motion mode, whose excitation is referred to as phonon, has a frequency $\omega_z /2 \pi =2.04~{\rm MHz}$. We cool the ion motion to its ground state by sideband cooling after Doppler cooling. We encode the TLS 
in the electronic levels $\ket{\downarrow} \equiv \ket{^{2}S_{1/2},m_j = 1/2}$ and $\ket{\uparrow} \equiv \ket{^{2}D_{5/2},m_j = 1/2}$. The electric quadrupole transition between these two levels can be driven by a 729 nm laser directly, corresponding to a resonant transition frequency denoted as $\omega_0$. Apply a set of 729 nm multi-component laser field to the ion, we obtain the system Hamiltonian Eq.\,(\ref{eq01}) in the basis  $\{\ket{g}=(\ket{\downarrow}-i\ket{\uparrow})/\sqrt{2}, \ket{e}=(\ket{\downarrow}+i\ket{\uparrow})/\sqrt{2}\}$\,\cite{supp}.
In the experiment, we use an arbitrary waveform generator to synthesize the signal applied on a single-pass acousto-optical modulator for the desired multi-component 729 nm laser field. The 729 nm laser is aligned along the trap axis with $45^\circ$ to the magnetic field.

\emph{Detection of the parity symmetry}.---Below we consider applying a probe pulse to drive the TLS  coherently. The probe pulse, denoted as $ \Omega \cos(\omega_L t) \sigma_x $, is composed of another set of 729 nm laser field\,\cite{supp}.
The total Hamiltonian becomes $H=H_s+\Omega\cos(\omega_L t)\sigma_x$. Under the condition of resonant driving $\Delta_{L\sigma}=\omega_L-\omega_{\sigma}=j\omega_a$ ($j=1,2,3\dots$), the model is first excited to the state $|0,e\rangle$ from the initial state $|0,g\rangle$ via the flip of the TLS, then the interaction between two subsystems induces the sideband transitions from $|0,e\rangle$ to $|j,e\rangle$\,\cite{Bin2021WL}. For the standard Rabi model, $\theta=\pi/2$, the presence of parity symmetry only allows the transition to $|2n,e\rangle$  ($j=2n$) from  $|0,e\rangle$, and the transition from $|0,e\rangle$ to $|2n-1,e\rangle$ is forbidden due to their opposite parities. This results in only the state $|2n,e\rangle$ being excited in this parity-symmetry-preserving model.  In the case $\theta\neq\pi/2$, the parity symmetry of the Hamiltonian Eq.\,(\ref{eq01}) is broken, allowing connections between two parity subspaces via parity-symmetry-breaking transitions. The transition from $|0,e\rangle$ is not confined to the odd parity subspace. Thus, the state $|n,e\rangle$ ($j=n$) with any $n$ can be excited in this parity-symmetry-breaking model. Figure\,1(b) shows the selection rules for sideband transitions starting from $|0,g\rangle$, decided by the system's parity symmetry. These phonon state excitations, governed by selection rules, are a powerful signature for identifying the presence or breaking of parity symmetry for extended QRM.

The experiment of probing the phonon excitation spectrum is conducted by recording the response of mean phonon number $\langle a^{\dag} a\rangle$ to different parameters $\{\theta, ~\omega_L\}$. The probing process starts at state $\ket{0,g}$ which is prepared from the initial state $\ket{0, \downarrow}$ by applying a carrier $\pi/2$ pulse driven by a 729 nm laser with frequency {$\omega_0$} and initial relative phase {$\pi$}.  
Then we switch on two sets of 729 nm lasers together, and the quantum states of the ion evolve under the laser field for a duration. Since the transition rate between $|0,g\rangle$ and $|j,e\rangle$ varies significantly with different values of  $j$ and $\theta$, we set the evolution duration accordingly to observe the distinct peaks of phonon excitation. 
To obtain the phonon number $\langle  a^{\dag} a\rangle $ after evolution, we utilize a standard method for extracting the phonon distribution from the blue sideband evolution\,\cite{Wineland1998}. This is done after resetting the state to $\ket{ \downarrow}$ using a series of optical pumping pulses. 

\begin{figure}
\includegraphics[width=8.7cm]{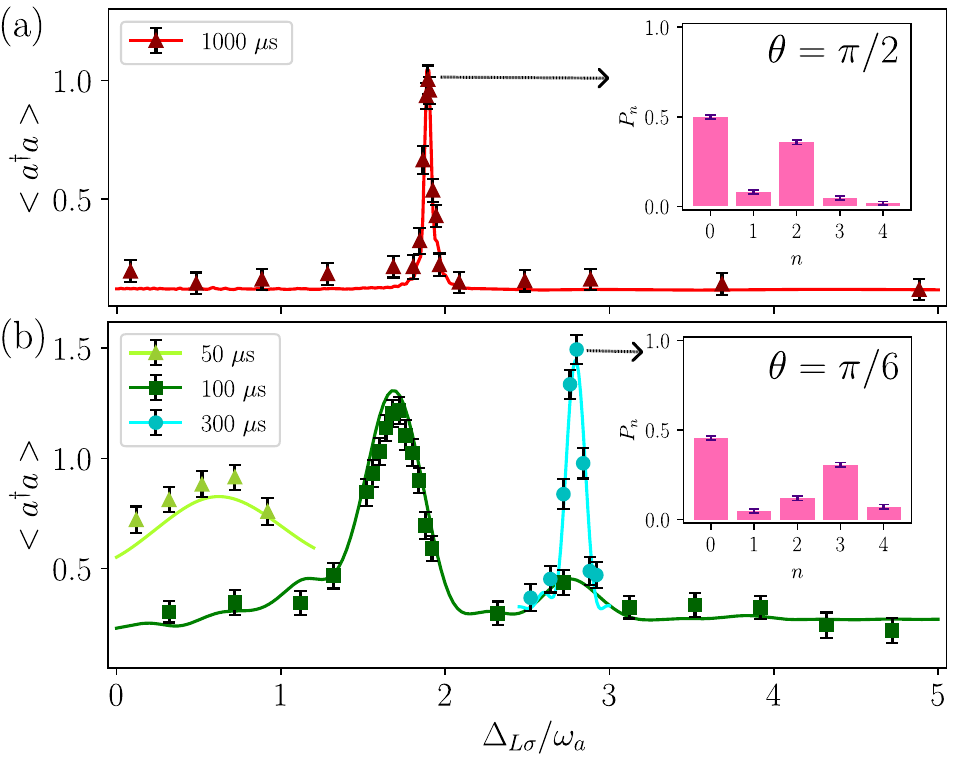}\\
\caption{Phonon excitation spectrum versus $\Delta_{L\sigma}/\omega_a$ for (a) $\theta=\pi/2$ and (b) $\theta=\pi/6$. The solid lines correspond to numerical simulations, while the experimental data is marked by triangles, squares, and circles. The experimental results are obtained by fitting the evolution of the blue sideband transition after a $1000 {\rm ~\mu s}$ probing pulse, with error bars indicating the fitting uncertainty. Inset: the phonon state distribution at the corresponding $\Delta_{L\sigma}/\omega_a$. In panel (b), triangles, squares, and circles correspond to the probing pulse with $50 {\rm ~\mu s}$, $100 {\rm ~\mu s}$, and $300 {\rm ~\mu s}$, respectively. System parameters are: $\lambda/\omega_a = 0.27$, $\omega_a/2 \pi = 25 {\rm ~kHz}$, (a) $\omega_{\sigma}/\omega_a = 4.32$, $\Omega/\omega_a = 0.95$, and (b) $\omega_{\sigma}/\omega_a = 6.08$, $\Omega/\omega_a = 0.93$.}
\label{fig2}
\end{figure}

The experimental results for two cases of $\theta$ are shown in Fig.\,\ref{fig2}, where we consider the ultra-strong coupling regime with $\lambda /\omega_a = 0.27$. In the standard Rabi model, $\theta=\pi/2$, Fig.\,2(a) displays a phonon excitation peak at the resonance condition  $\Delta_{L\sigma}\approx2\omega_a$ but not at $\Delta_{L\sigma}\approx1\omega_a,3\omega_a$. This peak mainly corresponds to the excitation of the state $|2,e\rangle$, as confirmed by phonon occupations shown in the inset of Fig.\,\ref{fig2}(a). This clearly demonstrates that the presence of parity symmetry allows only the excitation of states with even numbers of phonons. Although the theoretical model predicts phonon excitation could also occur at resonance $\Delta_{L\sigma}\approx4\omega_a$, neither numerical simulations nor experimental results show this at limited probe pulse duration due to system dissipation even under strong driving, which hampers transitions from a ground state to a higher-order state\,\cite{Bin2021WL, Munoz2014VT, Bin2020LL}. For the case of parity-symmetry-breaking, we choose $\theta=\pi/6$, where three phonon excitation peaks are visible at the resonance points $\Delta_{L\sigma}\approx1\omega_a,2\omega_a,3\omega_a$, corresponding mainly to the excitations of the states $|1,e\rangle, |2,e\rangle, |3,e\rangle$, respectively, as depicted in Fig.\,\ref{fig2}(b).  This indicates that the excitation of states with any number of phonons is possible when the parity symmetry is broken.  Note that all peaks in Fig.\,\ref{fig2} showing the resonance positions experience deviations from ideal resonances of $\Delta_{L\sigma}\approx j\omega_a$, primarily due to the dressing of the TLS caused by strong driving, which shifts the $j$-phonon resonance positions\,\cite{Mollow1969, Munoz2014VT, Carreno2017VL, Bin2020LL}. Instead, the $j$-phonon resonance position is recalculated as $\Delta_{L\sigma}=\sqrt{j^2 \omega_a^2-\Omega^2}$\,\cite{supp}.

\begin{figure*}
\includegraphics[width=17.4cm]{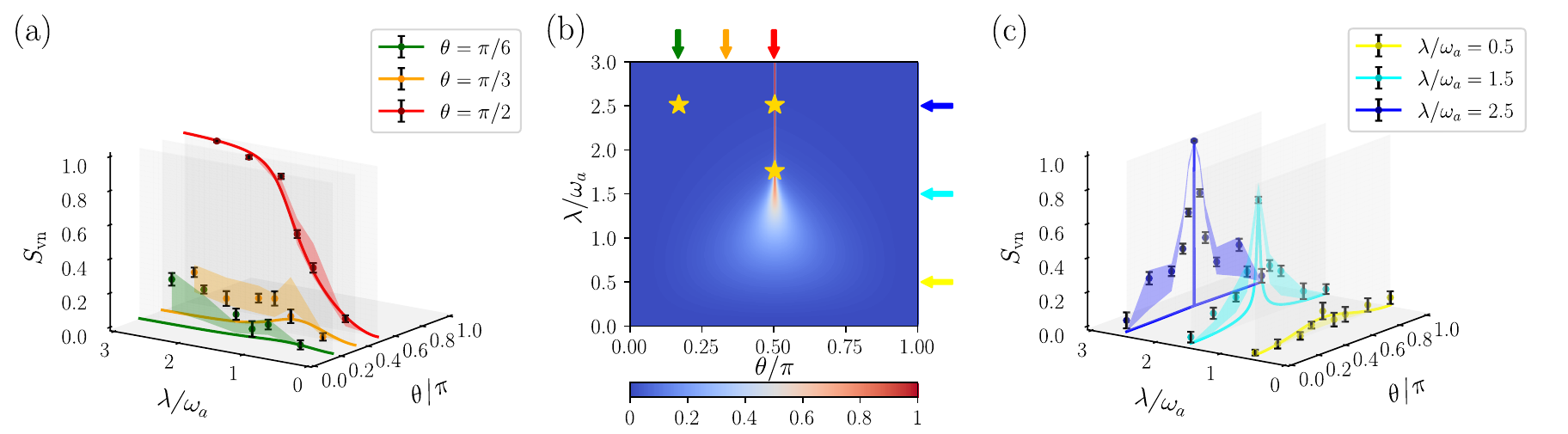}\\
\caption{Von Neumann entropy of the TLS. (a,\,c) The solid lines and dots correspond to the results obtained by numerical simulation and experiment, respectively. Error bars represent one standard deviation, and shaded regions show the numerical simulation results with shot-to-shot noise. (b) Von Neumann entropy versus $\lambda/\omega_a$ and $\theta/\pi$, obtained via numerical simulation. The arrows around the frame indicate the parameter sets for the results in (a,\,c), while the stars indicate the parameter set used in Fig.\,\ref{fig4}. Only for case $\theta = \pi/2$ in (a), $\omega_{\sigma}/\omega_a = 4.0$\,\cite{supp}. For other cases in (a-c), $\omega_{\sigma}/\omega_a = 4.5$. All cases take $\omega_a /2 \pi = 2.82 {\rm ~kHz}$.}
\label{fig3}
\end{figure*}

\emph{Parity-symmetry-protected quantum phenomena}.---Next, we focus on the influence of parity symmetry on the ground state. The ground state of $H_s$ exhibits significant differences in scenarios with and without parity symmetry, such as quasiprobability distributions in the phase space of the bosonic mode and entanglement between the TLS and the phonon mode. We should notice that parity symmetry is only conserved when $\theta=\pi/2$. Even small deviations from this value can lead to parity symmetry breaking, resulting in changes to quantum phenomena related to the ground state. The increase of $\lambda$ over critical value results in strong ground-state entanglement and quantum superposition, that does not occur in the parity-symmetry-breaking case. To observe such parity-symmetry-protected quantum phenomena, we adiabatically prepare the ground state of $H_s$ and calculate the von Neumann entropy of the TLS based on its state tomography results\,\cite{supp, Lv2018AL}. Additionally, we reconstruct the state of the phonon mode and confirm that a negative part appears in its Wigner distributions when parity symmetry is present.

The entanglement of a coupled system can, to some extent, be characterized by the von Neumann entropy of a subsystem. To assess the entanglement of two subsystems in QRM-namely, the TLS and phonon mode-we perform standard quantum state tomography\,\cite{Nilsen} for the TL\textcolor{blue}{S} after tracing out the phonon mode from the ground state. Then we calculate the von Neumann entropy $S_{\mathrm{vn}}$, defined as $S_{\mathrm{vn}} = -\mathrm{Tr}(\rho_{\sigma} \ln \rho_{\sigma})$, according to the results of the quantum state tomography $\rho_{\sigma}$. Figure\,\ref{fig3}(a) shows the experimental results for three cases of $\theta$ at different coupling strengths $\lambda$, ranging from $0.5\,\omega_a$ to $2.5\,\omega_a$. The solid lines and shaded bands represent the $S_{\mathrm{vn}}$ of the reduced density matrix $\rho_{\sigma}$ for ideal and noise-added ground states, respectively. The experimental data align with the prediction for the noise-added ground state case but show some deviations from the ideal ground state case, caused by typical decoherence in the trapped ion and additional shot-to-shot fluctuations of laser intensity\,\cite{supp}. The bifurcation of $S_{\mathrm{vn}}$ versus $\lambda$ for different values of $\theta$ demonstrates that the parity symmetry of the system plays a crucial role in the occurrence of the ground state entanglement. This parity-symmetry-controlled phenomenon is further illustrated in Fig.\,\ref{fig3}(c). A distinct peak of $S_{\mathrm{vn}}$  appears at $\theta=\pi/2$; here, the deviation near $\theta = \pi/2$ between the experimental data and results in the ideal ground state case can be attributed to a decline in fidelity $\mathcal{F}$, in addition to the same reasons as for Fig.\,\ref{fig3}(a). These experimental results take the parameters along the virtual lines indicated by arrows with the same color in Fig.\,\ref{fig3}(b), where the diagram depicts the $S_{\mathrm{vn}}$ of $\rho_{\sigma}$ for ideal ground states by direct diagonalization of $H_s$. The peak near $\theta = \pi/2$ sharpens as $\lambda$ increases. Extrapolating to the limit of strong coupling strength, $S_{\mathrm{vn}}$ will become an obvious signal, nearly adopting two discrete values $\{0,~1\}$  to reflect the breaking and presence of parity symmetry of the system.

To show the Wigner distributions, we reconstruct the reduced density matrix $\rho_a$ by directly measuring the characteristic function after tracing out the TLS from the ground state\,\cite{supp}. The real and imaginary components of the characteristic function are measured separately, and the Wigner distribution $\mathcal{W}(\gamma)$ is derived through a two-dimensional Fourier transform of the characteristic function\,\cite{Fluhmann}. Figure\,\ref{fig4} shows the Wigner distribution of $\rho_a$ for different parameters $\{\theta, ~\lambda / \omega_a\}$ marked by yellow stars in Fig.\,\ref{fig3}(b). At $\lambda / \omega_a = 1.75$ and $\theta = \pi / 2$, the Wigner distribution of $\rho_a$ in Fig.\,\ref{fig4}(a) displays two separated peaks with an interference fringe between them, confirming the existence of quantum superposition in the ground state across the critical point of the ground state phase transition in the parity-symmetry-conserving case. Increasing the coupling strength to $\lambda / \omega_a = 2.5$, the peaks move further apart in Fig.\,\ref{fig4}(b). However, the denser interference fringe predicted theoretically is not clearly visible mainly due to the limited resolution of the Wigner distribution based on the available data. For comparison, the Wigner distribution of $\rho_a$ at $\theta = \pi /6$ and $\lambda / \omega_a = 2.5$ in Fig.\,\ref{fig4}(c) shows only a single peak without interference fringes, where the phonon state $\rho_a$  behaves like a coherent state without superposition. The quantum superposition of the phonon mode of the ground state is another manifestation of parity symmetry, especially in the deep strong coupling regime. Figures\,\ref{fig4}(d-f) depict corresponding results for the ideal ground state. The deviation observed in experimental results can be attributed to decoherence and experimental errors, including the imperfect preparation of the zero phonon state after cooling.

\begin{figure}
\includegraphics[width=8.7cm]{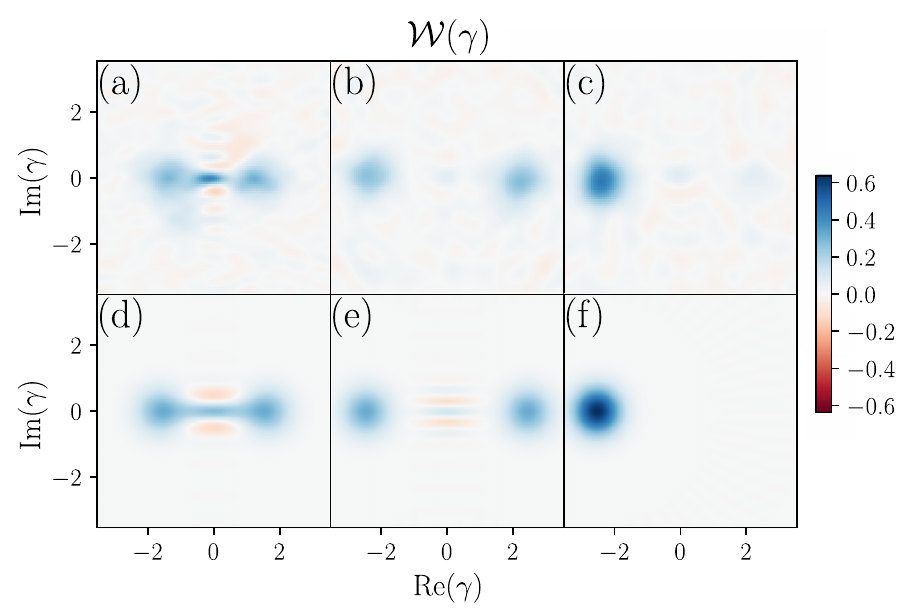}\\
\caption{Wigner distribution of the phonon mode. (a-c) These diagrams show the Wigner distribution reconstructed through a two-dimensional Fourier transform of the experimentally measured characteristic function. (d-f) Theoretical Wigner distribution of the phonon mode traced from the ideal ground state given by direct diagonalization of Hamiltonian. Parameters $\{\theta, ~\lambda / \omega_a\}$ take $\{\pi / 2,  1.75\}$ for (a,\,d), $\{\pi / 2,  2.50\}$ for (b,\,e), and $\{\pi / 6,  2.50\}$ for (c,\,f). All cases take $\omega_{\sigma}/\omega_a = 4.5$ and $\omega_a / 2 \pi = 2.82 {\rm ~kHz}$.}
\label{fig4}
\end{figure}

\emph{Conclusion}.--- In summary, we explore the parity symmetry in extended QRM experimentally based on a trapped ion system. A method for directly probing the parity symmetry is demonstrated. By observing the excitation spectrum of the phonon, we can identify the parity-symmetry-conserving transitions and parity-symmetry-breaking ones. This method can be used in other physical platforms as long as an additional driving term is applied. Furthermore, we observe the quantum phenomena strongly related to parity symmetry over large parameter regimes thanks to the versatility of trapped ions. 
It's particularly interesting that parity symmetry protects these quantum phenomena, including strong ground-state entanglement and quantum superposition, due to their potential applications in fields like quantum sensing and coherence enhancement. Our work verifies the fundamental results predicted by quantum theory and opens a door for the experimental exploration of symmetry-controlled quantum phenomena.

\begin{acknowledgments}
The USTC team acknowledge support from the National Natural Science Foundation of China (grant number 92165206) and Innovation Program for Quantum Science and Technology (Grant No. 2021ZD0301603). X.L. is supported by the National Science Fund for Distinguished Young Scholars of China (Grant No.\,12425502), the National Key Research and Development Program of China (Grant No.\,2021YFA1400700) and the Fundamental Research Funds for the Central Universities (Grant No.\,2024BRA001). Q.B. is supported by the National Natural Science Foundation of China  (Grant No.\,12205109) and Natural Science Foundation of Sichuan Province of China  (Grant No.\,25NSFSC1895). 
\end{acknowledgments}

\onecolumngrid
\clearpage
\setcounter{equation}{0}
\setcounter{figure}{0}
\setcounter{table}{0}
\setcounter{page}{6}
\setcounter{section}{0}
\makeatletter
\renewcommand{\theequation}{S\arabic{equation}}
\renewcommand{\thefigure}{S\arabic{figure}}
\renewcommand{\bibnumfmt}[1]{[S#1]}
\renewcommand{\citenumfont}[1]{S#1}

\renewcommand{\thefigure}{S\arabic{figure}}
\setcounter{figure}{0}
\renewcommand{\theequation}{S\arabic{equation}}
\setcounter{equation}{0}
\renewcommand{\thetable}{S\arabic{table}}
\setcounter{table}{0}

\begin{center}
        \textbf{Supplemental Material for ``Experimental observation of parity-symmetry-protected phenomena in the quantum Rabi model with a trapped ion"}
\end{center}

\title{Supplemental Material for ``Experimental observation of parity-symmetry-protected phenomena in the quantum Rabi model with a trapped ion"}
\date{\today}

\title{Supplemental Material for ``Experimental observation of parity-symmetry-protected phenomena in the quantum Rabi model with a trapped ion"}

\date{\today}
\maketitle

\section{I. Derivation of effective Hamiltonian}\label{section1}

The Hamiltonian of the ion interacted with multi-component laser field is described by ($\hbar=1$)
\begin{align}\label{eqI01}
H_m = \frac{\omega_0}{2} \sigma_z+\omega_z a^\dag a + \sum_m \frac{\Omega_m}{2} \{\sigma^{\dag} e^{i [ \eta (a + a^{\dag})- (\omega_m t + \varphi_m)]} + h.c.\}, ~m = 1, 2, 3... ~, 
\end{align}
where $\omega_0$ is the transition frequency of two levels, $\omega_z$ is the frequency of axial phonon mode (see main text) and $\{\Omega_m,~\omega_m ,~\varphi_m\}$ are Rabi rate, laser frequency and phase of individual component of laser field. We apply a unitary transformation 
\begin{align}\label{eqI02}
U = e^{-i[(\omega_0 - \omega_{\sigma})\sigma_z /2 + (\omega_z - \omega_a) a^\dag a] t}
\end{align}
with tunable parameters $\{\omega_{\sigma}, \omega_a\}$ to Hamiltonian $H_m$, and expand it to second order in $\eta$. Then we get
\begin{align}\label{eqI03}
H' = \frac{\omega_{\sigma}}{2} \sigma_z + \omega_a a^\dag a + \sum_m \frac{\Omega_m}{2} \{
\sigma^{\dag} [1 + i \eta (a e^{i(-\omega_z + \omega_a)t} + a^{\dag} e^{i(\omega_z - \omega_a)t})] e^{i[( \omega_0 - \omega_{\sigma} - \omega_m )t - \varphi_m]}
+ h.c.\}.
\end{align}
We take laser field with $\{\Omega_1,~\Omega_2,~\Omega_3,~\omega_1,~\omega_2,~\omega_3,~\varphi_1,~\varphi_2,~\varphi_3\} = \{ \omega_0 -\omega_z + \omega_a, ~\omega_0 +\omega_z - \omega_a, ~\omega_0, ~2\lambda/ \eta, ~2\lambda/ \eta, ~\omega_q, ~\pi -\theta, ~ \pi -\theta, ~\pi/2 \}$ as mentioned in main text, and apply rotating wave approximation. Then Eq.\,(\ref{eqI03}) becomes
\begin{align}\label{eqI04}
H'' = \frac{\omega_{\sigma}}{2} \sigma_y + \omega_a a^\dag a + \lambda (\cos \theta\sigma_y\!-\!\sin \theta \sigma_x)(a + a^{\dag}).
\end{align}
Considering additional laser components $\{\Omega_4,~\Omega_5,~\omega_4,~\omega_5,~\varphi_4,~\varphi_5\} = \{ \omega_0 -\omega_L, ~\omega_0 +\omega_L, ~\Omega, ~\Omega, ~0, ~0\}$. Then Eq.\,(\ref{eqI04}) get a new term $\Omega\cos(\omega_L t)\sigma_x$ and become $H'''$. This Hamiltonian can be mapped to 
\begin{align}\label{eqI05}
\begin{split}
H = & ~ S H''' S^{\dag}\\
= & ~\frac{\omega_{\sigma}}{2} \sigma_z + \omega_a a^\dag a + \lambda (\cos \theta\sigma_z\!-\!\sin \theta \sigma_x)(a + a^{\dag}) + \Omega\cos(\omega_L t)\sigma_x.
\end{split}
\end{align}
by transformation $S=((-1,i), (i,-1))/\sqrt{2}$ which redefine the ground state $\ket{g}$ and excited state $\ket{e}$ of two-level system in $H$ as $(\ket{\downarrow}-i\ket{\uparrow})/\sqrt{2}$ and $(\ket{\downarrow}+i\ket{\uparrow})/\sqrt{2}$ respectively (the global phase was omitted here).

\section{II. Derivation of multiquanta resonances conditions}\label{section2}
 
When considering the probe pulse driving on the two-level system, we remind the total Hamiltonian Eq.\,(\ref{eqI05}). In the parameter regime $\Omega\ll\omega_{\sigma}$, the influence of the driving probe pulse on the energy structure of the Hamiltonian can be safely ignored, the ideal $j$-phonon resonance condition is obtained by $\Delta_{\sigma L}=\omega_{\sigma}-\omega_L=-j\omega_a$.
 
In the large pumping regime $\Omega\sim\omega_{\sigma}$, the strong driving probe pulse can dress the atom, and the two-level system forms new eigenstates that are a quantum superposition of the bare states $\{|e\rangle, |g\rangle\}$. The subsystem Hamiltonian for the strongly driven two-level system at a frame rotating with the probe pulse frequency $\omega_L$ can be approximately given by 
\begin{align}\label{eqII02}
H_{\sigma}=&\frac{1}{2}(\omega_{\sigma}-\omega_L)\sigma_z+\frac{1}{2}\Omega(\sigma^\dag +\sigma ),
 \end{align}
with the eigenvalues
\begin{align}\label{eqII03}
E_{|\pm\rangle}=\pm\frac{1}{2}\sqrt{(\omega_{\sigma}-\omega_L)^2+\Omega^2},
 \end{align}
and corresponding eigenstates
\begin{align}\label{eqII04}
&|+\rangle=c_+|g\rangle+c_-|e\rangle,~~~ |-\rangle=c_-|g\rangle-c_+|e\rangle,
 \end{align}
where
\begin{align}\label{eqII05}
&c_{\pm}=\sqrt{\frac{\Omega^2/2}{(\omega_{\sigma}-\omega_{L})^2+\Omega^2\pm(\omega_{\sigma}-\omega_{L})\sqrt{(\omega_{\sigma}-\omega_{L})^2+\Omega^2}}}
 \end{align}
 and $c_+^2+c_-^2=1$. Together with the phonon modes and ignoring the influences of the ultrastrong interaction on the energy structure, the eigenvalues become
\begin{align}\label{eqII06}
E_{|j,\pm\rangle}=j\omega_a\pm\frac{1}{2}\sqrt{(\omega_{\sigma}-\omega_L)^2+\Omega^2}.
\end{align}
By setting $E_{|0,+\rangle}=E_{|j,-\rangle}$, i.e., $\sqrt{(\omega_{\sigma}-\omega_L)^2+\Omega^2}=j\omega_a$, we obtain the $j$-phonon resonance condition 
$\Delta_{\sigma L}=\omega_{\sigma}-\omega_L=-\sqrt{j^2\omega_a^2-\Omega^2}$. Therefore, compared to the ideal $j$-phonon resonance conditions, here we have a resonance shift $|\delta|=\sqrt{j^2\omega_a^2-\Omega^2}-j\omega_a$ caused by the strong driving on the two-level system, as shown in Fig.\,2 of the main text.

\section{III. Details of adiabatic ground state preparation}\label{section3}

Preceding the experimental execution of adiabatic preparation, we design a specialized path for different combinations of parameters $\{ \theta, ~ \lambda \}$ with the help of numerical computation by {\footnotesize QUTIP}\,\cite{Johansson2012}. Generally, we start from the state $\ket{0,\downarrow}$, and apply a $\pi / 2$ pulse to initially prepare $\ket{0,g}$. The adiabatic preparation then proceeds by adjusting the parameter $\omega_a$ according to the formula $\omega_a(t-t_0) = (\omega_{\mathrm{max}}-\omega_{\mathrm{tar}})e^{-(t-t_0)/\tau}+\omega_{\mathrm{tar}}, ~t_0 \le t \le t_{\mathrm{tot}}$, where $t_{\mathrm{tot}}$ is total evolution time, $t_0$ is the time span from the start of the $\pi/2$ pulse to the start of the scan. Similar to\,\cite{Lv2018AL}, we calculate the fidelity $\mathcal{F}$ between the instantaneous ground state $\rho_{\mathrm{in}}$ of the Hamiltonian $H_s$ and the adiabatically evolved state $\rho_{\mathrm{ad}}$, defined as $\mathcal{F} = \operatorname{Tr} \sqrt{\sqrt{\rho_{\mathrm{in}}} \rho_{\mathrm{ad}} \sqrt{\rho_{\mathrm{in}}}}$. For simplicity, we fix $\omega_{\mathrm{max}} = 2 \pi \times 150.00~{\rm kHz}$ and $\omega_{\mathrm{tar}} \simeq 2 \pi \times 2.82 ~{\rm kHz}$ for all adiabatic paths,  and we search for appropriate values of $\{\tau, ~t_{\mathrm{tot}}\}$ for each path. Specifically for paths in Fig.\,3(a) of main text, we ensure that the numerical value of  $\mathcal{F} > 0.98$ at the end of the adiabatic preparation process in the absence of noise (listed in Table.\,\ref{tableS1}). We should notice that we takes $\omega_{\sigma}/\omega_a = 4.0$ only for case $\theta = \pi/2$ to increase $\mathcal{F}$. Without loss of generality, we compare the von Neumann entropy of the two-level system of the ideal ground state for $\omega_{\sigma}/\omega_a = 4.0$ with that for $\omega_{\sigma}/\omega_a = 4.5$ as shown in Fig.\,S1(a). For adiabatic paths with the same parameter $\lambda$ in Fig.\,3(c) of main text, we set the same values of $\{\tau, ~t_{\mathrm{tot}}\}$ as the ones in Table.\,\ref{tableS1} sine the decoherence in experiment limits the value $t_{\mathrm{tot}}$.  We show the curve of fidelity $\mathcal{F}$ versus $\theta$ for different $\lambda$ in Fig.\,S1(b). We can observe a decline in $\mathcal{F}$ near $\theta = \pi /2 $ for the paths we chose here, which contributes to part of the deviation between the experimental and theoretically ideal results.

\begin{table*}[b]
 \begin{center}
  \caption{Parameters for different adiabatic paths and corresponding fidelity at the end of the adiabatic preparation process. The fidelity is calculated numerically according to effective Hamiltonian Eq.\,(\ref{eqI05}) without noise. For cases $\theta = \pi/2$, $\omega_{\sigma}/\omega_a = 4.0$. For cases $\theta = \pi/6$ and $\pi/3$, $\omega_{\sigma}/\omega_a = 4.5$.}
  \label{tableS1}
  \begin{tabular}{|c|c|c|c|c|c|c|c|}
   \hline
   Parameters & \diagbox{$\theta$}{$\lambda/\omega_a$} & 0.50 & 1.00 & 1.25 & 1.50 & 2.00 & 2.50 \\
   \hline
   \multirow{3}{*}{$\tau / {\rm \mu s}$} & $\pi/6$ & \multirow{3}{*}{40} & \multirow{3}{*}{60} & \multirow{3}{*}{70} & \multirow{3}{*}{80} & \multirow{3}{*}{100} & \multirow{3}{*}{80}\\
   \cline{2-2}
   & $\pi/3$ & & & & & & \\
   \cline{2-2}
   & $\pi/2$ & & & & & & \\
   \hline
   \multirow{3}{*}{$t_{\mathrm{tot}} / {\rm \mu s}$} & $\pi/6$ & \multirow{3}{*}{100} & \multirow{3}{*}{300} & 400 & \multirow{3}{*}{600} & \multirow{3}{*}{700} & \multirow{3}{*}{600}\\
   \cline{2-2}
   \cline{5-5}
   & $\pi/3$ & & & 500 & & & \\
   \cline{2-2}
   \cline{5-5}
   & $\pi/2$ & & & 400 & & & \\
   \hline
   \multirow{3}{*}{$\mathcal{F}$} & $\pi/6$ & 1.000 & 0.986 & 0.982 & 0.998 & 0.999 & 0.995\\
   \cline{2-8}
   & $\pi/3$ & 1.000 & 0.987 & 0.983 & 0.992 & 0.999 & 0.994\\
   \cline{2-8}
   & $\pi/2$ & 1.000 & 1.000 & 0.998 & 0.992 & 0.980 & 0.983\\
   \hline
  \end{tabular}
  \end{center}
\end{table*}

\begin{figure}
\includegraphics[width=17.4cm]{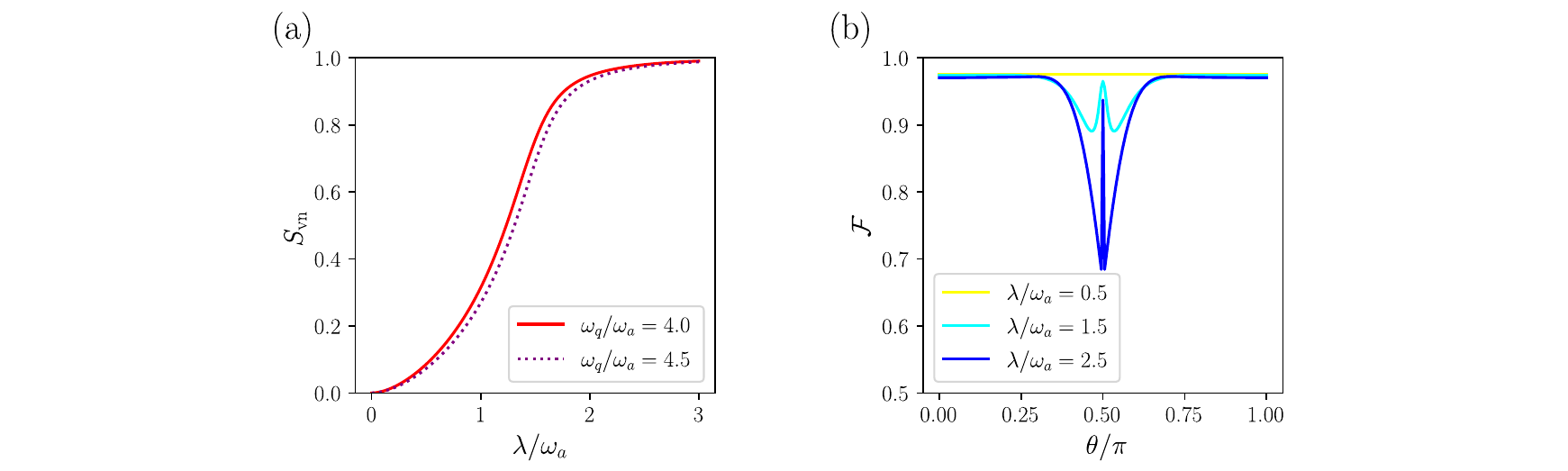}\\
\caption{Compare of von Neumann entropy of the two-level system and fidelity at the end of different adiabatic paths. (a) For cases $\theta = \pi/2$, the solid line and dotted line show similar results of the ideal von Neumann entropy of the two-level system given by direct diagonalization of Hamiltonian for $\omega_{\sigma}/\omega_a = 4.0$ and $4.5$. (b) The solid lines represent fidelity $\mathcal{F}$ versus $\theta$ for $\lambda=0.5,~1.5$ and $2.5$.}
\label{fig4}
\end{figure}

\section{IV. shot-to-shot noise in simulation}\label{section4}
The evolution quantum state can be simulated numerically using package {\footnotesize QUTIP}\,\cite{Johansson2012} by solving the Lindblad master equation 
\begin{align}\label{eqIV01}
\dot{\rho}(t)=-\frac{i}{\hbar}[H(t), \rho(t)]+\sum_n \frac{1}{2}\left[2 C_n \rho(t) C_n^{\dagger}-\rho(t) C_n^{\dagger} C_n-C_n^{\dagger} C_n \rho(t)\right]
 \end{align}
where $\rho$ is density matrix of system and $C_n$ are collapse operators. For trapped ions, we usually consider three kinds of decoherence mechanisms. More specifically, $C_1=\sqrt{1/\tau_1} \sigma_z$ corresponds to dephasing of the two-level system with decoherence time $\tau_1 = 1.5~{\rm ms}$, $C_2=\sqrt{1/\tau_2} a^\dag a $ corresponds to dephasing of phonon mode with decoherence time $\tau_2 = 8.0~{\rm ms}$, $C_3=\sqrt{\gamma(n_{th})} a^\dag $ and $C_4=\sqrt{\gamma(n_{th}+1)} a$ contribute together to motional heating of phonon mode with heating rate $\gamma(n_{th}) \approx \gamma(n_{th}+1) = 60~{\rm s^{-1}}$. In addition to these decoherence mechanisms, the evolution of the ion's state in our experimental platform is also affected by shot-to-shot noise such as laser intensity fluctuation, laser frequency fluctuation, and imperfect cooling.

In the experiment for probing parity symmetry, we detect the phonon distribution at the end of evolution which is not sensitive to the shot-to-shot noise mentioned above. In consequence, we only consider the decoherence mechanisms as collapse operators and set the initial phonon state as the thermal state with mean phonon number $\bar{n}=0.05$ for simulation in Fig.\,2. As for experiments related to von Neumann entropy, we take dephasing and motional heating of phonon mode as collapse operators in the Lindblad master equation and sample the laser intensity and frequency related by AC Stark shift as the normal distribution with parameters measured independently. These parameters include the coefficient of variation of laser intensity (0.0226) and the maximum of AC Stark shift caused by laser field (5.459 ${\rm kHz}$). Here we emphasize that the stationary change of frequency $\omega_0$ caused by AC Stark shift is calibrated for laser field with different components. Moreover, we also sample the $\bar{n}$ of the initial phonon state as the unitary distribution in a range $[0, ~ 0.1]$ to simulate the effect of imperfect cooling.
The simulation results are plotted as bands in Fig.\,3(a,\,c).

\section{V. Von Neumann entropy}\label{section5}
Von Neumann entropy is a crucial concept in quantum information theory and is also usually used to depict properties of quantum states\,\cite{Nielsen}. We give the entanglement information between two QRM subsystems using von Neumann entropy $S_{\mathrm{vn}}$ of the two-level system after tracing out phonon mode. The trace operation used here for the two-level system differs from the spin reset used for phonon states as mentioned in the main text. This operation is carried out spontaneously before applying analytic carrier pulses for quantum state tomography because short carrier pulses are insensitive to the phonon state if the mean phonon number $\left< a^{\dag} a\right>$ is not high. For all ground states prepared adiabatically in this experiment, we numerically verify that the influence of carrier pulses imposed by the phonon distribution can be neglected. Moreover, the experimental results of the quantum state tomography of the two-level system are transformed into $S_{\mathrm{vn}}$ by the Monte Carlo method according to the definition of $S_{\mathrm{vn}}$.

\section{VI. Wigner function}\label{section6}
The Wigner distribution $\mathcal{W}(\gamma)$ fully describes a phonon state in the position momentum phase space $(x, p)$. It can be obtained by the two-dimensional Fourier transform of the characteristic function $\chi(\beta)$:
\begin{align}\label{eqVI01}
\mathcal{W}(\gamma)=\frac{1}{\pi^2} \int \chi(\beta) e^{\gamma \beta^*-\gamma^* \beta} d^2 \beta,
 \end{align}
where $\chi(\beta)=\langle{\mathcal{D}}(\beta)\rangle$ with ${\mathcal{D}}(\beta)=e^{\beta {a}^{\dagger}-\beta^* {a}}$ being displacement operator and angle brackets $\langle \cdot \rangle$ denoting expectation value on phonon state. Here $a$ ($a^\dag$) represents the annihilation (creation) operator mentioned in the main text and the real (imaginary) parts of $\gamma$ and $\beta$ represent the coordinate of $x$ ($p$). As shown in Fig.\,S2, we directly measure characteristic functions' real and imaginary parts using the method described in work\,\cite{Fluhmann} and then combine them to reconstruct the Wigner distribution in Fig.\,4. Each diagram in Fig.\,S2 is collected in a grid with 41$\times$41 size, and each point represents data with 300 times repetition. 
We need to mention that theoretically predicted imaginary parts of characteristic functions vanish when $\theta$ takes $\pi /2$. However, the imperfect preparation of the zero phonon state after cooling causes shallow patterns to appear in Fig.\,S2(d,\,e). This subsequently deforms the Wigner distribution slightly in Fig.\,4. 

\begin{figure*}
\includegraphics[width=17.4cm]{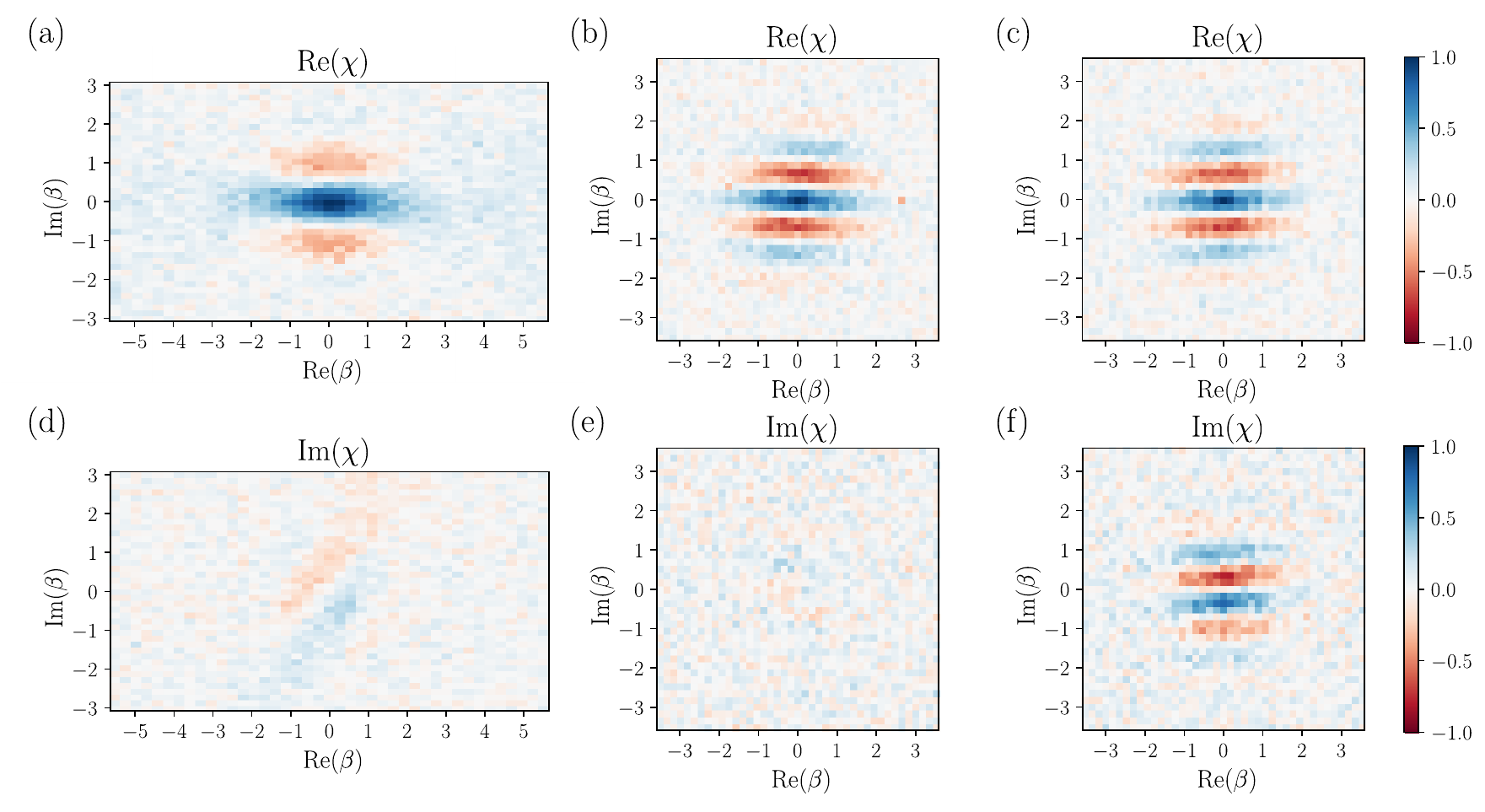}\\
\caption{Characteristic functions of phonon mode. (a-c) Measured characteristic functions' real parts of different phonon states traced from the experimentally prepared ground state. (d-f) Measured characteristic functions' imaginary parts of corresponding phonon states in (a-c). Parameters $\{\theta, ~\lambda / \omega_a\}$ take $\{\pi / 2,  1.75\}$ for (a,\,d), $\{\pi / 2,  2.5\}$ for (b,\,e), and $\{\pi / 6,  2.5\}$ for (c,\,f). Parameters $\{\tau, ~t_{\mathrm{tot}}\}/ {\rm \mu s}$ take $\{100, 700\}$ for (a,\,d), $\{80,  600\}$ for (b,\,c,\,e\,f).}
\label{fig3}
\end{figure*}

\end{document}